# Hydrogen-modified inverse Spin Hall Effect in palladium-cobalt bi-layer films

S. Watt[1], R. Cong[1,2], C. Lueng[1], M. Sushruth[1], P.J. Metaxas[1], and M. Kostylev[1*]

[1]*School of Physics and Astrophysics, University of Western Australia, 6009 Crawley, WA, Australia*

[2]*University of Science and Technology of China, Hefei, China*

*Corresponding author, email: mikhail.kostylev@uwa.edu.au

Abstract: The influence of hydrogen gas absorption by the Pd layer of bi-layered films containing Pd and Co layers on the inverse Spin Hall Effect (iSHE) in the material is measured. iSHE is driven by ferromagnetic resonance in the cobalt layer. In these conditions, the iSHE is seen as a d.c. voltage across the Pd layer. In the presence of hydrogen gas the iSHE peak shifts downwards in the applied field together with the ferromagnetic resonance absorption peak for the material. In parallel, an increase in the iSHE peak height and in its width is observed. Our analysis suggests that these observations can potentially be explained as a reduction in the spin Hall angle for the Palladium layer in the presence of hydrogen gas.

Palladium (Pd)/cobalt (Co) bi-layer films demonstrate a host of important magnetic and spintronic effects. Firstly, a strong perpendicular magnetic anisotropy (PMA) is induced at the interface between the two layers [1]. Also these films are characterized by a Spin Hall Effect. The inverse Spin Hall voltage across the Pd layer can be observed during driven ferromagnetic resonance (FMR) in the Co layer [2]. A strong spin-pumping effect [3] across the interface underlies the strong inverse Spin Hall Effect (iSHE).

Unique for palladium is its affinity to hydrogen gas ($H_2$) at room temperature and normal atmospheric pressure. It has been demonstrated that absorption of $H_2$ affects the effective field of interface PMA [4-6]. This results in a downward shift (in field) of the in-plane FMR peak for the cobalt layer [7]. The shift is consistent with a hydrogen-induced decrease in PMA.

It is possible to exploit the $H_2$-induced FMR peak shift to measure $H_2$ concentration in the environment [8-9]. With the introduction of hydrogen fuel cells in the automotive industry, the demand for highly stable and sensitive $H_2$ sensors has risen dramatically and there is an unsatisfied requirement for safer, more sensitive detection methods [10]. Many approaches to developing $H_2$ sensors make use of Palladium's affinity to $H_2$. Sensors exploiting FMR-based probing of the interfacial PMA in Pd/Co bi-layer films have an important advantage: demonstrated capability of $H_2$ concentration measurements in a very wide concentration range, including near 100% hydrogen concentration [9] (a concentration regime where most alternative approaches fail).

In the present paper we explore the effect of $H_2$ on the iSHE response of Co/Pd films when their Co layers are driven to FMR. We find that the peak of iSHE voltage across the Pd layer shifts downwards in applied field in the presence of the gas. The work demonstrates the potential for FMR-based $H_2$ sensing using a device-generated *dc* output, here the iSHE voltage. In contrast, in our previous studies, the output signal of the magnetic gas sensor was a *microwave* voltage [8]. Using the iSHE voltage eliminates the need to convert the originally microwave-frequency sensor output into a low-frequency signal that is suited to detection with simple electronics. The observed $H_2$-induced iSHE peak shift is accompanied by a decrease in the peak linewidth and an increase in its height. Our analysis suggests that $H_2$-induced changes to the iSHE peak height may originate from an impact of $H_2$ on the Spin Hall angle for the Pd layer.

Cobalt/palladium bilayer films were deposited onto silicon substrates using DC magnetron sputtering at an argon working pressure of 8 mTorr. The system base pressure prior to sputtering was $10^{-7}$ Torr. Each sample consists of a 10 nm layer of cobalt with a capping palladium layer of varying thickness (4.4 nm - 15 nm).

Leads were attached to opposite sample edges using a conductive epoxy. The samples were mounted on a polystyrene holder and placed in the centre of a microwave cavity where they were subject to an in-plane external magnetic field (Fig. 1). The microwave power source was tuned to the cavity resonance at approximately 9.5 GHz and the external field swept to induce FMR in the cobalt layer. The iSHE-induced voltage across the sample length perpendicular to the in-plane field was measured using a nanovoltmeter.

The microwave cavity also has built-in modulation coils. With these coils, a small ac. modulating field of frequency 20.1 kHz is applied to the sample parallel to the static magnetic field. The microwave signal reflected from the cavity is incident onto a microwave diode which is used to obtain field-resolved FMR spectra. To do this, the rectified voltage from the diode's output port is fed into a lock-in amplifier locked to the frequency of the modulation field ("Field Modulated FMR"). Accordingly, the lock-in output scales as the first derivative of the resonance absorption signal with respect to the applied field ("differential FMR trace").

Each sample is measured first in nitrogen gas ($N_2$) atmosphere and then exposed to a $H_2$-containing atmosphere. We use 3% $H_2$ diluted in 97% of $N_2$ (note that the flammability threshold for $H_2$ in air is 4% [11]).

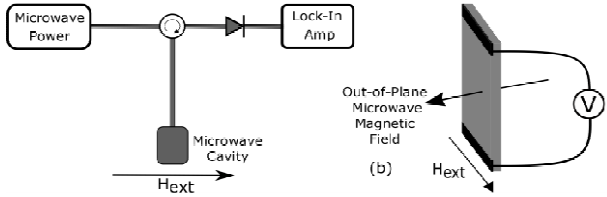

**Fig. 1. (a)** Simplified microwave cavity FMR configuration. **(b)** Voltage across the sample measured perpendicular to external field and incident microwave field via leads attached with conductive epoxy.

Typical FMR and iSHE responses are shown in Fig. 2. One notices that the differential FMR absorption trace is very close to perfectly anti-symmetric. However, the iSHE voltage signal is not perfectly symmetric, although one would expect symmetric iSHE peaks for our experimental geometry. Based on [12] we speculate that the voltage signal has two contributions; the symmetric iSHE and an anti-symmetric anisotropic magnetoresistance. The latter has dependence on orientation and, actually, should be negligible when the static magnetisation vector is aligned perfectly perpendicular to the measured voltage. Its presence in our data possibly suggests that there is a small, unintended misalignment of the two.

Upon exposure of the sample to the $H_2$-containing gas mixture (referred to as "3%$H_2$" in the following), a shift to lower fields of both iSHE and FMR peaks is evident. The iSHE peak is fitted with a combination of Lorentzian and anti-Lorentzian line-shapes and the FMR peak with a combination of the first derivatives of those shapes (Eqs.(13) and (8a) in Appendix). From these fits, the resonance position and linewidth are extracted.

The fits demonstrate a very good agreement of the FMR and iSHE peak positions for all the samples. This is consistent with the established understanding that the spin pumping effect by the precessing magnetization in the Co layer underlies the iSHE response (see e.g. theories in [14,12]).

Based on those theories, one would also expect the FMR and ISHE traces to share the same linewidth. However, we found a discrepancy between the linewidths extracted. The explanation for this discrepancy is beyond the scope of this paper and below we concentrate on the changes in the traces upon hydrogenation. These changes are illustrated in Figs. 3 and 4. They can be summarized as follows.

We find a consistent decrease in the resonance field upon hydrogenation (Figs. 2 and 3(a)). The peak shift to lower fields is consistent with previous studies [7-9, 15-17] and is related to a decrease in the effective field of the interface PMA. The $H_2$-induced iSHE peak shift is the main finding of this work. It demonstrates the possibility of reading out the state of a resonance-based magnetic $H_2$ sensor directly at dc using the iSHE. Interestingly, a minimum decrease in the FMR field is clearly observed around 8 nm of Palladium. As expected, the ISHE peak shifts follow a similar trend, being a direct consequence of the resonating spins.

We also find that $H_2$ absorption decreases the linewidths (Fig. 3(b)) of both FMR and iSHE peaks. This observation is consistent with our previous studies in which we also saw a decrease in the FMR peak linewidth [7,8]. Above we mentioned that the linewidths of the iSHE and FMR peaks appeared to be different. However, the ratio between the linewidths obtained in 3%$H_2$ and $N_2$ were approximately the same for iSHE and FMR (Fig. 4 (b)).

In our previous work, we consistently observed an increase in the FMR peak height upon $H_2$ absorption by the stack [7,8,16]. The same trend is seen in Figs. 2(a). However, from Fig.2(b) one notices that the change in the iSHE peak height is smaller. Fig. 4(c) illustrates this effect - it shows the ratio of resonance peak heights for the FMR and ISHE traces in 3%$H_2$ and $N_2$ atmospheres.

In addition to FMR measurements, we also took measurements of the film total resistance in $N_2$ and upon sample exposure to 3%$H_2$. We observed a decrease in the resistance upon hydrogenation. This is illustrated with Fig. 4(a), where the ratio between the resistances in 3%$H_2$ and nitrogen is displayed. On average, the decrease is 10% with respect to the virgin ($N_2$) state of the samples. Fig. 5 demonstrates the repeatability of the change in resistance during cycling between 3%$H_2$ and pure $N_2$. One sees that the change is repeatable starting from the second cycle, and that there is some irreversible component of the change taking place during the first cycle.

Let us now discuss these observations. Theoretically, the FMR absorption peak height is inversely proportional to the resonance linewidth $\Delta H$. Furthermore, the iSHE voltage peak height scales as $1/(\Delta H)^2$. The latter follows from Eq.(5) in [14] and both scalings are explained in Appendix.

One explanation for the observed height increase is thus the observed decrease in $\Delta H$. We may exclude from consideration this contribution to the total decrease in the height by dividing the ratio of peak heights by the ratio of the respective $\Delta H$ or $(\Delta H)^2$, depending on the trace type. Below we will call the result of this data manipulation "ratio of response amplitudes". It is explained in detail in Appendix and shown in Fig. 4(d). One sees that, all the iSHE points in this panel lie below the 1.0 level. Hence, hydrogenation noticeably decreases the Pd layer ability to generate iSHE in response to magnetization precession in the Co layer. This implies that the increase in the iSHE peak height seen in Fig. 2 is mainly due to a decrease in the FMR linewidth. A smaller $\Delta H$ means smaller magnetic losses in the Co layer; the decrease in the loss parameter leads to a larger magnetisation precession angle for the same amplitude of the microwave driving field producing a stronger iSHE response.

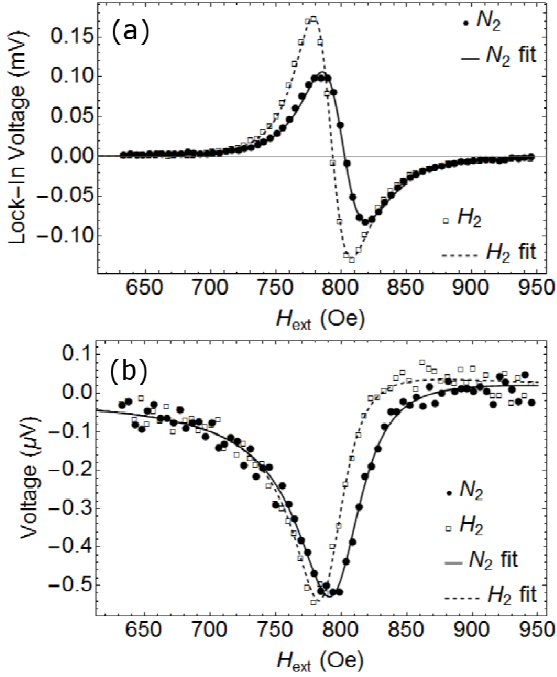

Fig. 2. (a) FMR signal and fit and (b) iSHE voltage and fit for Co(10 nm)/Pd(6.3 nm) sample in nitrogen (circles/solid lines) and 3%H2 (squares/dashed lines).

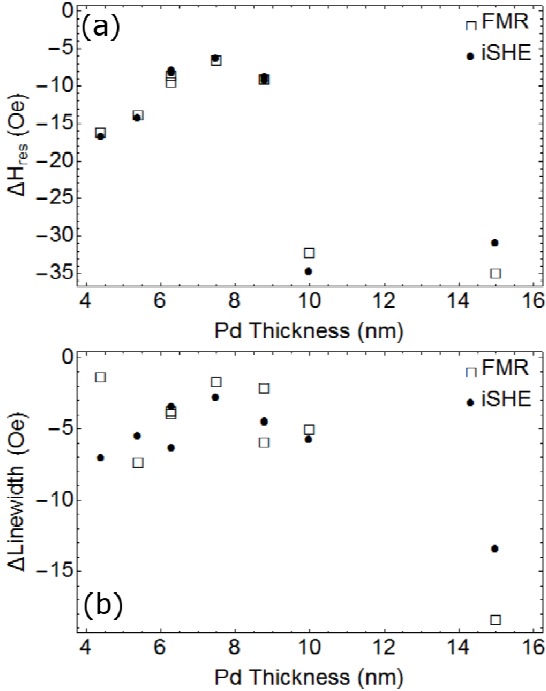

**Fig. 3.** (a) Shift in resonance position and (b) change in resonance linewidth upon hydrogenation measured via FMR (squares) and iSHE (circles).

One important observation from Fig. 4(d) is that the increase in the height of the FMR peaks is larger than one expects from a mere decrease in $\Delta H$. This is evidenced by most of the FMR points in Fig. 4(d) lying above the 1.0 level. Equation (10) in Appendix shows that a H2-induced change in the PMA strength may also lead to a change of the FMR peak height through a change in the strength of coupling of the dynamic magnetization to the driving field. However, numerical evaluation of this equation reveals that this contribution is negligible with respect to the total change in the height (see a numerical example in Appendix). Hence, there should be a different reason for the FMR peak height growth.

Above we mentioned that the iSHE response amplitude decreases in the presence of $H_2$ (Fig. 4(d)). From Eq.(12) in Appendix it follows that there may be a contribution to the decrease in the iSHE response amplitude which is related to the effect of the decrease in the PMA field leading to an increase of the coupling of dynamic magnetisation to the driving field. To account for this effect, Fig. 4(d) actually displays iSHE data multiplied by a correction coefficient given by Eq. (14) in Appendix. The coefficient has been evaluated for each sample separately based on the values of the effective saturation magnetisation extracted for the sample from the respective FMR peak positions for both atmospheres. However, the correction does not exceed 1% for any sample. Hence it has a negligible effect on the overall behaviour of the iSHE peak height, as the points in Fig. 4(d) lie well below the 1.0 level.

One way to explain the reduction in the iSHE signal is by assuming that the strength of spin pumping across the interface is reduced in the presence of $H_2$. A reduction in spin pumping should be seen as a decrease in $\Delta H$ [18]. The resonance line narrowing seen in the FMR traces (Fig. 4 (b)) is in agreement with this idea. However, our observation of the decrease in the samples' resistance (Fig. 4(a)) is in some contradiction with the assumption. The irreversible part of the resistance decrease may be related to the effect of hydrogen annealing of the sample [19]. However, the annealing cannot explain the reversible change. Usually Pd resistance increases in the presence of $H_2$ [20], however, the opposite effect of a reversible Pd resistance decrease upon hydrogen absorption is observed for thin layers and small $H_2$ concentrations and is related to Pd film growth through island formation with consequent island merging [21]. We suggest that the same explanation may apply to our films.

As follows from the expressions in [22], a decrease in the Pd layer resistance should result in an increase in the spin mixing conductance at the interface and hence improve the efficiency of spin pumping across it instead of reducing it. This suggests that the iSHE response amplitude decreases in the presence of $H_2$, because hydrogenation also affects other parameters upon which the amplitude depends. In Eq.(5) in [14], these parameters are grouped into the quantity $K$. Besides the spin-mixing conductance whose role we have discussed above, several other quantities enter the expression, including the Spin Hall angle (SHA). This suggests that the anticipated increase in the spin-mixing conductance due to the reduction in the Pd layer resistivity may be overcompensated by a $H_2$-induced decrease in SHA. SHA depends on the strength of spin-orbit coupling for the material, and, potentially, hydrogen incorporation into Pd lattice could influence the coupling.

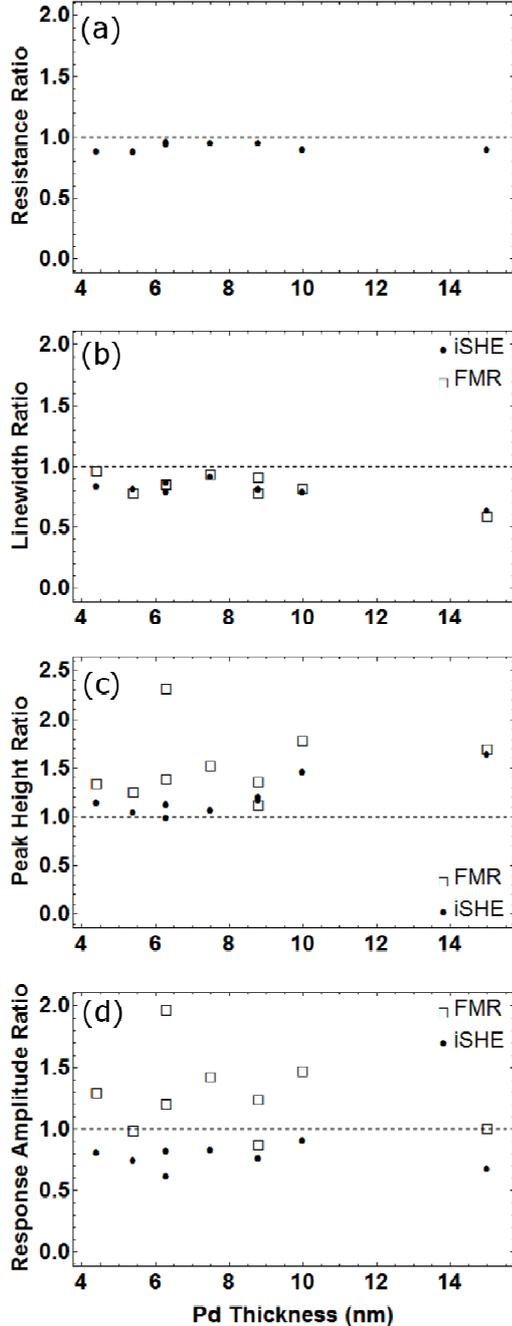

**Fig. 4.** (a) Ratio of Pd resistivities, (b) Ratio of iSHE (circles) and FMR (squares) linewidths, (c) Ratio of iSHE (circles) and FMR (squares) peak heights, and (d) Ratio of iSHE (circles) and FMR (squares) response amplitudes between nitrogen and 3%$H_2$ atmospheres. The values in Panel (d) have been corrected to account for the effect of the change in the PMA strength on the amplitudes.

In conclusion, in this work, we carried out investigation of the influence of $H_2$ absorption by the Pd layer of bi-layered films containing Pd and Co layers on iSHE in the material while it is driven by FMR in the cobalt layer. We found that in the presence of $H_2$ the iSHE peak shifts downwards in the applied field together with the FMR absorption peak for the material.

In parallel, an increase in the iSHE peak height and a decrease in the peak width were observed for most of samples. However, the change in the iSHE peak height is somewhat smaller than the change of the same parameter for the FMR peaks. Our analysis shows that this observation can potentially be explained as a reduction in the Spin Hall angle for the Palladium layer in the presence of $H_2$.

This finding opens up a pathway to design a magnetic-film based $H_2$ sensor whose output signal will be a dc voltage. This may represent an advantage with respect to previously suggested magnetic $H_2$ sensor concept based on a microwave frequency sensor output [8].

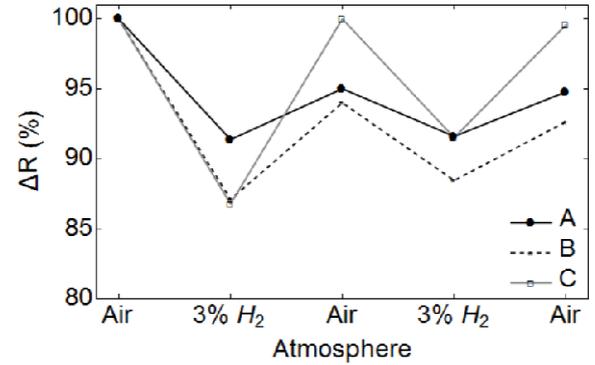

**Fig. 5.** Relative change in resistance of Co/Pd samples as a function of atmosphere for some of the studied samples. Sample A: Co/Pd[10nm], Samples B and C: Co/Pd[5.4nm].

Acknowledgement
The authors acknowledge the facilities, and the scientific and technical assistance of the Australian National Fabrication Facility at the Centre for Microscopy, Characterisation & Analysis, The University of Western Australia, a facility funded by the University, State and Commonwealth Governments. M.K. acknowledges his Senior Research Award from the University of Western Australia. P.J.M. acknowledges support from the Australian Research Council's Discovery Early Career Researcher Award Scheme (DE120100155).

Appendix

Equation (1) from Ref. [12] can be cast in the form as follows

$$j_s \propto 2\omega \text{Im}(\dot{m}_x \bar{m}_y), \quad (1)$$

where $j_s$ the time-averaged spin current across the interface between the ferromagnetic (cobalt) and the non-magnetic (palladium) layer, $\omega$ is microwave frequency, $\dot{\mathbf{m}} = (\dot{m}_x, \dot{m}_y)$ is the complex-valued amplitude of dynamic magnetization $\mathbf{m}(t) = \dot{\mathbf{m}}\exp(i\omega t)$ and the dash on top of $\dot{m}_y$ denotes complex conjugation. The spin current is polarized in the direction $z$ of the applied magnetic field $H$. The field and the axis $x$ lie in the film plane and the axis $y$ is along the normal to the film surface. For the conditions of the spatially uniform magnetization vector precession in an in-plane magnetized ferromagnetic layer, the linearized Landau-Lifshitz Equation can be cast in the vector-matrix form as follows:

$$\lambda \dot{\mathbf{m}} - \hat{C}\dot{\mathbf{m}} = \mathbf{f}, \qquad (2)$$

where $\lambda = i|\gamma|\mu_0 H/\omega$, $\gamma$ is the gyromagnetic ratio,

$$\hat{C} = \begin{bmatrix} 0 & -1 \\ 1 & -i|\gamma|\mu_0 M_{eff}/\omega \end{bmatrix}, \qquad (3)$$

the column vector $\mathbf{f} = (i|\gamma|\mu_0 M \dot{h}_s / \omega, 0)$, $\dot{\mathbf{h}}_s = (\dot{h}_s, 0)$ is the amplitude of the microwave driving field applied along the $x$ direction, $M_{eff} = M_s - H_{PMA}$ is the effective magnetization which represents the difference between the saturation magnetization for the ferromagnetic layer $M_s$ and the effective (bulk-like) field of the interface perpendicular anisotropy $H_{PMA}$. By expanding $\hat{C}$ in terms of its right-hand and left-hand eigen-vectors one arrives at a solution for $\dot{\mathbf{m}}$

$$\dot{\mathbf{m}} = \frac{\dot{h}_s}{H - (H_0 + i\Delta H)} \frac{M_s}{\Lambda} \dot{\boldsymbol{\varepsilon}}, \qquad (4)$$

where $H_0$ is the ferromagnetic resonance field for frequency $\omega$

$$H_0 = \sqrt{[\omega/(\mu_0|\gamma|)]^2 + (M_{eff}/2)^2} - M_{eff}/2, \qquad (5)$$

$$\Lambda = \sqrt{M_{eff}^2 + [2\omega/(\mu_0|\gamma|)]^2},$$

$\Delta H$ is the magnetic loss parameter for the cobalt layer, and $\dot{\boldsymbol{\varepsilon}}$ is a column-vector:

$$\dot{\boldsymbol{\varepsilon}} = \begin{bmatrix} i\frac{\Lambda + M_{eff}}{2\omega/(|\gamma|\mu_0)} \\ 1 \end{bmatrix}. \qquad (5a)$$

The loss parameter was introduced into Eq.(4) phenomenologically, by adding an imaginary part to the resonance field $H_0$. This procedure creates a complex resonance field $H_0 + i\Delta H$.

The energy absorbed by the material in the presence of a microwave magnetic field scales as $\bar{\mathbf{h}}_s \cdot \dot{\mathbf{m}}$. Hence the shape of the FMR absorption trace $A(H)$ is given by $\sqrt{\bar{\mathbf{h}}_s \cdot \dot{\mathbf{m}}}$. Then, the height of the maximum of the resonance line (the height of the resonance peak) which is located at $H = H_0$ scales as follows

$$|A(H_0)| \propto \frac{r}{\Delta H_s} |\dot{h}_s|, \qquad (6)$$

where the "response amplitude"

$$r = \frac{|\gamma|\mu_0 M_s}{\sqrt{2}\,\omega}\sqrt{1 + \frac{M_{eff}}{\Lambda}}. \qquad (7)$$

As both $A$ and $r$ are known to a constant, experimentally the response amplitude $\dot{r}^{(\exp)}$ is obtained by fitting the recorded FMR traces with an equation which follows from Eq.(4):

$$\dot{A}_{\exp}(H) = \frac{\dot{r}^{(\exp)}}{H - (H_0 + i\Delta H)} \qquad (8)$$

which implies that the height of the experimental peak $|\dot{A}_{\exp}(H_0)| = |\dot{r}^{(\exp)}/\Delta H|$.

As in our experiments we utilize the Field-Modulated FMR, the actual experimental traces have the shape given by

$$\mathrm{Re}(d\dot{A}_{\exp}(H)/dH)$$
$$= \frac{d}{dH}\left(\frac{(H-H_0)\,\mathrm{Re}(\dot{r}^{(\exp)})}{(H-H_0)^2 + \Delta H^2} - \frac{\Delta H\,\mathrm{Im}(\dot{r}^{(\exp)})}{(H-H_0)^2 + \Delta H^2}\right), \qquad (8a)$$

where $\dot{r}^{(\exp)}$ is allowed to take complex values in order to account for potential asymmetries of the FMR absorption lines.

Accordingly, the theoretical ratio of the FMR peak heights for $H_2$ and $N_2$ atmospheres is given by

$$\left|\dot{A}_{H_2}(H_{0H_2})/\dot{A}_{N_2}(H_{0N_2})\right|_{teor} = \frac{\Delta H_{N_2}}{\Delta H_{H_2}}\frac{r_{H_2}}{r_{N_2}}, \qquad (9a)$$

and the experimental one by

$$\left|\dot{A}_{H_2}(H_{0H_2})/\dot{A}_{N_2}(H_{0N_2})\right|_{\exp} = \frac{\Delta H_{N_2}}{\Delta H_{H_2}}\frac{|\dot{r}_{H_2}^{(\exp)}|}{|\dot{r}_{N_2}^{(\exp)}|}, \qquad (9b)$$

where $H_{0H_2}$ and $H_{0N_2}$ are the resonance fields for the $H_2$ and pure nitrogen atmospheres respectively. The theoretical ratio of the response amplitudes $r_{H_2}/r_{N_2}$ follows from Eq.(6):

$$\frac{r_{H_2}}{r_{N_2}} = \sqrt{\frac{1 + \dfrac{M_S - H_{PMA}^{H_2}}{\sqrt{\left(M_S - H_{PMA}^{H_2}\right)^2 + \left[2\omega/(|\gamma|\mu_0)\right]^2}}}{1 + \dfrac{M_S - H_{PMA}^{N_2}}{\sqrt{\left(M_S - H_{PMA}^{N_2}\right)^2 + \left[2\omega/(|\gamma|\mu_0)\right]^2}}}}. \qquad (10)$$

In the presence of hydrogen gas both $H_{PMA}$ and the resonance linewidth $\Delta H$ decrease. Hence one may expect an increase in the peak height $|\dot{A}(H_0)|$ because of the decrease in $\Delta H$. On top of this, one may expect a slight correction, also upwards, due to a slight increase in *r*. Physically, the latter correction originates from a slight increase in the elipticity of magnetization precession due to a decrease in $H_{PMA}$ (see Eq.(5a)). Therefore, while comparing the heights of the experimental peaks for the two atmospheres it is useful to also look at $\left| \dot{r}_{H_2}^{(exp)} / \dot{r}_{N_2}^{(exp)} \right| = \left| \dot{A}_{H_2}(H_{0H_2}) / \dot{A}_{N_2}(H_{0N_2}) \right| \left( \Delta H_{N_2} / \Delta H_{H_2} \right)$ and compare it with the theoretical ratio of the response amplitudes $r_{H_2} / r_{N_2}$. If one finds that $\left| \dot{r}_{H_2}^{(exp)} / \dot{r}_{N_2}^{(exp)} \right| / (r_{H_2} / r_{N_2}) \neq 1$ then there should be a contribution to the total change in the height of the experimental peak originating from a process other than the trivial change in the elipticity of precession.

Similarly, from Eq.(1) one finds that

$$|j_s(H)| \propto \frac{\rho}{(H-H_0)^2 + \Delta H^2} h_s^2, \quad (11)$$

where the iSHE response amplitude

$$\rho = \frac{|\gamma| \mu_0 M_S^2 [M_{eff} + \Lambda]}{2\omega \Lambda^2}. \quad (12)$$

From Eq.(11) one sees that the spin current should grow in the presence of hydrogen gas because of the decrease in $\Delta H$. The growth should be stronger than for the FMR absorption peak height, because of the second-power dependence on the loss parameter. The increase in $M_{eff}$ will compensate this growth, at least partly, as follows from the expression for $\rho$ (12).

Experimentally, the iSHE response amplitude and linewidth can be obtained by fitting the iSHE traces by an equation similar to Eq.(8):

$$A_{exp}(H) = \frac{\rho_s^{(exp)}}{(H-H_0)^2 + \Delta H^2} + \frac{(H-H_0)\rho_a^{(exp)}}{(H-H_0)^2 + \Delta H^2}. \quad (13)$$

This allows correcting for potential asymmetry of the measured voltage peak by assuming that the experimental iSHE response amplitude is given by the amplitude $\rho_s^{(exp)}$ of the symmetric component of the function $A_{exp}(H)$ only.

Similarly to the analysis of the FMR absorption peak height, it is useful to analyze the behaviour of $\rho_s^{(exp)}$ as a function of the sample environment. Again, in doing so, one can eliminate from consideration the effect of the change in $\rho$ caused by the change in elipticity of precession. This is done by looking at the ratio $\left( \rho_{sH_2}^{(exp)} / \rho_{sN_2}^{(exp)} \right) / (\rho_{H_2} / \rho_{N_2})$, where the correction coefficient $\rho_{H_2} / \rho_{N_2}$ follows from Eq.(12)

$$\rho_{H_2} / \rho_{N_2}$$
$$= \frac{M_S - H_{PMA}^{H_2} + \sqrt{(M_S - H_{PMA}^{H_2})^2 + [2\omega/(|\gamma|\mu_0)]^2}}{M_S - H_{PMA}^{N_2} + \sqrt{(M_S - H_{PMA}^{N_2})^2 + [2\omega/(|\gamma|\mu_0)]^2}} \quad (14)$$
$$\times \frac{(M_S - H_{PMA}^{N_2})^2 + [2\omega/(|\gamma|\mu_0)]^2}{(M_S - H_{PMA}^{H_2})^2 + [2\omega/(|\gamma|\mu_0)]^2}$$

Let us now use the formalism above in order to analyze the experimental traces for the sample with the 5.4nm-thick Pd layer. Assuming a value of gyromagnetic ratio of 0.3 MHz/mT for the cobalt layer, from the position of the FMR peak for the virgin state of the sample (815 Oe=81.5mT) we extract a value of 1.15T for the effective magnetization $\mu_0 M_{eff}$. In the presence of hydrogen gas the peak shifts downwards by 11.2 Oe (1.12 mT). This translates into an increase in the effective magnetization of 17.9 mT. Then using Eq.(10) we obtain $r_{H_2} / r_{N_2} = 1.001$. The difference of this result from 1 is negligible. Hence one may neglect the effect of the change in the elipticity of precession on the height of the FMR absorption peak. Similarly, using Eq.(14) we obtain $\rho_{H_2} / \rho_{N_2} = 0.99$. One sees that there is a 1% decrease in the height of the iSHE peak. Accordingly, the iSHE data shown in Fig. 4(d) in the paper have been obtained by dividing the original $\rho_{H_2}^{(exp)} / \rho_{N_2}^{(exp)}$ by the respective $\rho_{H_2} / \rho_{N_2}$.